\documentclass{kluwer}    

\newdisplay{guess}{Conjecture}

\begin{document}                                                                                   
\begin{article}

\def\Ha{H$\alpha$}
\def\hi{{\sc H\thinspace i}}
\def\hii{{\sc H\thinspace ii}}
\def\Zsol{Z_\odot}
\newcommand{\Lx}{$L_{\rm x}$}
\newcommand{\kms}{\rm km\,s^{-1}}
\newcommand{\ergs}{\rm erg\,s^{-1}}
\newcommand{\teff}{$T_{\rm eff}$}
\newcommand{\etal}{{et\thinspace al.}~}
\def\spose#1{\hbox to 0pt{#1\hss}}
\def\lta{\mathrel{\spose{\lower 3pt\hbox{$\mathchar"218$}}
     \raise 2.0pt\hbox{$\mathchar"13C$}}}
\def\gta{\mathrel{\spose{\lower 3pt\hbox{$\mathchar"218$}}
     \raise 2.0pt\hbox{$\mathchar"13E$}}}
\newcommand{\Dt}{\spose{\raise 1.5ex\hbox{\hskip5pt$\mathchar"201$}}}    
\def\Mdot{\Dt{M}}

\begin{opening}         
\title{Superbubble Activity in Star-Forming Galaxies}
\author{M. S. \surname{Oey}}  
\runningauthor{M. S. Oey}
\runningtitle{Superbubble Activity in Galaxies}
\institute{Lowell Observatory, 1400 W. Mars Hill Rd., Flagstaff, AZ\ \ \ 86001,
	USA}
\date{November 30, 2002; {\it Ap\&SS,} in press}

\begin{abstract}

Mechanical feedback from massive stars, primarily from
supernovae, can dominate ISM structuring and phase balance, thereby
profoundly affecting galactic evolutionary processes.
Our understanding of mechanical feedback is based on the adiabatic,
wind-driven bubble model, applied on size scales ranging over three
decades. 
Tests of the model, and our consequent understanding of feedback, are
reviewed.  While the model is broadly successful, critical unknowns
still prevent a comprehensive understanding of the consequences of
feedback. 

\end{abstract}

\end{opening}           

\section{Introduction}  

Mechanical feedback from massive stars is a dominant driver
of evolutionary processes in galaxies, and takes place on
scales ranging from individual wind-driven bubbles to galactic superwinds.  
Our understanding of the feedback process is based on the standard
evolutionary model for stellar wind- and 
supernova-driven bubbles (Pikel'ner 1968; Weaver
{\etal}1977):  hot (log $T/{\rm K}\sim6-7$), low-density ($n\sim 0.01\
\rm cm^{-3}$) gas is generated within a double-shock structure, and
the pressure 
of this hot gas, chemically enriched by the stellar products, drives
the growth of the thin, radiatively-cooled shell.  In the adiabatic
model, energy loss from the hot gas is negligible,
yielding simple analytic expressions for the shell radius $R$ and
expansion velocity $v$ as a function of time $t$:
\begin{eqnarray}\label{eqAD}
R & \propto & (L/n)^{1/5}\ t^{3/5} \quad , \nonumber \\
v & \propto & (L/n)^{1/5}\ t^{-2/5} \quad .
\end{eqnarray}
For a stellar wind-driven bubble, the mechanical luminosity 
$L=1/2\ \Dt{M} v_\infty^2$, where $\Dt{M}$ and $v_\infty$ are the wind 
mass-loss rate and terminal velocity, respectively.  For OB
associations, supernovae (SNe) quickly dominate over winds, in which case
$L=N_* E_{51}/t_e$ (e.g., Mac Low \& McCray 1988), where $N_*$ and
$E_{51}$ are the total number of SNe and SN energy,
respectively, and $t_e$ is the time over which the SNe occur.

The fate of the interior hot gas is crucial to the phase balance
and enrichment of the interstellar and intergalactic media.  Is
stellar feedback indeed the source of the diffuse, hot gas in the ISM?
Do galactic superwinds from starbursts eject metal-enriched gas from
galaxies?  How does mechanical feedback affect the structure of the
ISM and porosity, for example, to ionizing radiation?  Do superbubbles
trigger renewed star formation?  

Determining the relevance of the standard, adiabatic model for shell
evolution is clearly critical in answering these fundamental
questions.  Several tests can be applied:  1) Comparing the observed
vs predicted dynamics of individual bubbles and superbubbles; 2)
Comparing the observed vs predicted statistical properties of 
superbubble populations, for example size, velocity, and energy
distributions; 3) Identifying spatial correlations of superbubbles
with the progenitor OB associations and their relics; and 4) Testing
the observed vs predicted dynamics and properties of galactic
superwinds.  The last is presently more difficult and the subject of
entire reviews in its own right (e.g., Heckman 2002).  Therefore I
will discuss here only the first three tests.

\section{Individual shell systems}

\subsection{Single star bubbles}

Few studies exist of single star bubbles from isolated OB stars.
Oey \& Massey (1994) examined two nebular examples in M33, and
spectroscopically classified the parent O stars.  The inferred
stellar masses and ages implied wind parameters that were consistent
with the observed sizes and shell ages predicted by the adiabatic
model.  However, the parameters were loosely constrained.
\hi\ shells with radii of several tens
of pc have been identified as wind-blown bubbles around a number of
Galactic O and Of stars (Cappa \& Benaglia 1998; Benaglia \& Cappa
1999).  These are largely consistent with the standard model, and
probe a specific subset of fairly evolved stars with old shells that
have essentially stopped expanding.

Studies of Wolf-Rayet ring nebulae suggest shells that are too small,
equivalent to an overestimate 
in $L/n$ by an order of magnitude (e.g., Treffers \& Chu 1982;
Garc\'\i a-Segura \& Mac Low 1995; Drissen {\etal}1995).  However, the
progenitor star produces several wind phases, including both fast and
slow winds, with extreme changes in $L$.  Their cumulative effect 
on shell morphology is complex and poorly understood.
It is therefore unsurprising to find
significant discrepancies between the predictions and observations of
shell parameters.  Hence, more studies of the simpler,
single OB star bubbles are needed.

\subsection{Superbubbles}

Superbubbles around OB associations are more prominent than
single-star bubbles, and thus have been studied more extensively.
Soft X-ray emission has been detected within many
objects, which is qualitatively consistent with the
adiabatic evolution model.  Two classes of X-ray
emission have been identified:  objects with X-ray luminosity \Lx\ in
excess of the model's prediction (Chu \& Mac Low 1990; Wang
\& Helfand 1991), and objects that remain undetected in X-rays (Chu
{\etal}1995).  The X-ray--bright objects are thought to be
overluminous because of SNR impacts on the shell walls.  Upper limits
on the X-ray--dim objects remain consistent with \Lx\ predicted by the
adiabatic model.  It will thus be of great interest to determine \Lx\
for these objects with {\it XMM-Newton} or {\it Chandra}.  
Also, an interface region between the hot gas and cooler shells should
generate intermediate temperatures and ions.  Chu {\etal}(1994)
searched a number sightlines 
through LMC superbubbles and confirmed the existence of
C\thinspace{\sc iv} and Si\thinspace{\sc iv} absorption in all cases.

A stringent test of the adiabatic model is to compare the predicted
and observed shell kinematics in cases where the input mechanical power and
other parameters are well-constrained.  This was carried out for
eight, young, wind-dominated LMC superbubbles by Oey \& Massey (1995), Oey
(1996), and Oey \& Smedley (1998).  The predicted growth rate for the
shells was higher than implied by their observed $R$ and $v$,
equivalent to an overestimate in $L/n$ by 
an order of magnitude.  However, even after adjusting $L/n$ in the
models, over half the objects still showed observed expansion
velocities that were typically a factor of two higher than predicted
for the given $R$.  Similar discrepancies were reported for
Galactic objects by Saken {\etal}(1992) and Brown {\etal}(1995).

SNR impacts on the shell wall are the favored explanation 
for the high-velocity shells, since these also exhibit the anomalously
high X-ray emission and elevated [S II]/\Ha\ ratios.  However,
a sudden drop in the ambient density can
induce a ``mini-blowout'' with shell kinematics that can easily
reproduce the anomalous velocities (Oey \& Smedley 1998; Mac Low
{\etal}1998; Silich \& Franco 1999).  Indeed, were it not for the
X-ray and nebular diagnostics, it would be impossible
to distinguish the shell acceleration mechanism from the
kinematics alone. 

Thus we see that the ambient properties are critical in
determining the shell evolution.  An underestimate in $n$ could, for
example, contribute to the growth rate discrepancy described above,
that is seen in all the objects.  To clarify the ambient
gas distribution, Oey {\etal}(2001) mapped the \hi\ distribution
within a $\sim40^\prime$ radius of three nebular LMC superbubbles at
30$^{\prime\prime}$ resolution.  The results show neutral environments 
that vary to an extreme, despite morphologically similar optical
nebulae.  It is therefore essentially impossible to infer
properties of the ambient material without direct,
multi-wavelength observations.

Another vital parameter for shell evolution is the ambient pressure,
$P_0$, which determines whether and when the superbubble growth becomes
pressure-confined.  While $P_0$ is usually unimportant in young,
high-pressure superbubbles like the nebular objects mentioned above,
it is of vital importance in the mid- to late-stage evolution.  It may
also be relevant in high-pressure, ionized environments like dense
star-forming regions (e.g., Garc\'\i a-Segura \& Franco 1996).  
Ultimately, $P_0$ determines the final size of the shells,
and conditions relative to blowout.  The uniformity and distribution
of $P_0$ in the multiphase ISM is therefore especially relevant to a
global understanding of superbubbles in galaxies (see below). 

Finally, if the hot gas within superbubbles does not blow out
and merge into the hot, ionized medium (HIM), it
is likely that the objects will cool and depart from energy conservation.
Indeed, whether and how the hot interior cools has long been a major
question for superbubble evolution and the fate of the hot gas.
Thermal conduction at the interface between the cool shell wall and
hot gas should cause a high rate of mass-loading into the interior.
The evaporated shell material dominates the
mass in the hot region, which could be further supplemented by
evaporation and ablation from small clouds overrun by the expanding
shocks (e.g., Cowie \& McKee 1977; McKee {\etal}1984; Arthur \&
Henney 1996).  If the interior density is 
sufficiently increased, radiative cooling will dominate, and the
shells will no longer grow adiabatically.  In addition, Silich
{\etal}(2001) point out the importance of enhanced metallicity
in the superbubble interiors, caused by the stellar and SN yields.
Preliminary investigation for individual objects by Silich \& Oey
(2001) shows enhancement in \Lx\ by almost an order of
magnitude for low-metallicity ($Z=0.05\ \Zsol$) objects.  
This increase in the cooling rate could facilitate a transition from
adiabatic to momentum-conserving evolution. 

\subsection{Supergiant shells}

The very largest \hi\ shells, having sizes of order 1 kpc,
emphasize some of the problems with the mechanical feedback model, and
also highlight possible alternative shell-creating mechanisms.

The existence of infalling high-velocity clouds (HVCs) suggests that
the impact of these objects could be an important contributor to 
supergiant shell populations.  This suggestion is consistent with
galactic fountain models for disk galaxies (e.g., Shapiro \& Field
1976), which are ultimately also powered by mechanical feedback in the
disk.  A number of hydrodynamical simulations of infalling HVCs
confirm that these impacts result in shell-like structures (e.g.,
Tenorio-Tagle {\etal}1986; Rand \& Stone 1996; Santill\'an
{\etal}1999). 

In addition, tidal effects, which dominate energetics and structure
formation at the largest length scales, could also create \hi\ 
features that resemble shells.  Note that many SN-driven shells will
not exhibit expansion velocities if they have become pressure-confined
by the ambient medium, thus a lack of observed expansion velocities
cannot distinguish between the feedback model and other models.  It
has been suggested that some of the largest holes in, e.g., M33 are
simply morphologically-suggestive inter-arm regions (Deul \& den Hartog
1990).  The same may be true of the giant hole identified by
de Blok \& Walter (2000) in NGC 6822.  Simple self-gravity effects
have also produced shell- and hole-like structures in numerical
simulations (Wada {\etal}2000), although morphologically these
structures appear more filamentary than the observations.

While such alternative mechanisms for creating shell-like structures
undoubtedly contribute to the supergiant shell population, 
the conventional mechanical feedback model nevertheless also
appears to apply in many situations.  
Meaburn's (1980) LMC-4 is a well-known example that is unambiguously
linked to Shapley's Constellation~III, a large, extended complex
of young stars.  Kim {\etal}(1999) are able to identify an evolutionary 
sequence for supergiant shells in the LMC, based on the relative loci of
\Ha\ and \hi\ emission.  In addition, Lee \& Irwin (1997) considered
formation mechanisms for supergiant shells in the edge-on SBc galaxy
NGC 3044.  They found no evidence of HVCs, and since the galaxy is
isolated, tidal interactions are also unable to explain the supergiant
shells.  They therefore conclude that the active star formation seen
in NGC 3044 most likely explains its supergiant shell structures.

Thus, probably both mechanical feedback and other mechanisms
form supergiant shell structures.  Presumably different processes
dominate under different circumstances, and these
remain to be understood.

\section{Statistical properties of superbubble populations}

The statistical properties of the superbubble populations offer
another test of the standard evolutionary model for the shells.
Oey \& Clarke (1997) derived expressions for the differential size
distribution $N(R)\ dR$ of superbubbles in a uniform ISM, using the
analytic expressions for adiabatic evolution (equation~\ref{eqAD}).
We considered a power-law mechanical luminosity function for the
parent OB associations, 
\begin{equation}\label{eqMLF}
\phi(L)\ dL = A L^{-\beta}\ dL \quad, 
\end{equation}
with $\beta\simeq 2$, which is robustly associated with the \hii\ region
luminosity function (e.g., Kennicutt {\etal}1989; Oey \& Clarke 1998a). 
The superbubble growth is taken to be pressure-confined when the
interior pressure $P_i = P_0$.  Star formation is assumed to be coeval
within each OB association, with SNe therefore exploding over a period
$t_e = 40$ Myr, the lifetime of the lowest-mass SN progenitors.  
For constant star-formation rate $\psi$ and power-law $\phi(L)$, we
found that:
\begin{equation}\label{eqcaseC}
N(R) \propto R^{1-2\beta} \quad ,
\end{equation}
effectively yielding $N(R)\propto R^{-3}$ for $\beta = 2$.  
Oey \& Clarke (1997) also derive $N(R)$ for 
other combinations of $\psi$ and $\phi(L)$.

This result agrees well with the \hi\ shell catalog for the Small
Magellanic Cloud (SMC) compiled by Staveley-Smith {\etal}(1997).  This
is by far the most complete sample of \hi\ shells obtained for any
galaxy, as evidenced by the fact that the relative number counts of
\hii\ regions and \hi\ shells are in excellent agreement with their
relative life expectancies.  For shells having $R\geq100$ pc, the
fitted power-law slope $\alpha=1-2\beta$ is $2.7\pm 0.6$, in excellent
agreement with the general prediction of $\alpha=3$.    

We note that different models for ISM structure
yield different predictions for $N(R)$.  For example, Stanimirovi\'c
{\etal}(1999) suggest a possible fractal structure for the neutral ISM.  From
the same \hi\ dataset of the SMC, they find a fractal dimension implying
a size distribution for \hi\ holes of $\alpha=3.5$.  It is difficult to
empirically differentiate this from our model, having $\alpha=3$; but it
is worth noting that the predictions are intrinsically different.  

However, the superbubble size distribution presently is not a
sensitive test in determining whether or not the objects evolve
adiabatically.  If all the internal energy is radiated away,
the objects are predicted to follow the momentum-conserving law given
by Steigman {\etal}(1975):
\begin{equation}
R\propto (L/nv_\infty)^{1/4}\ t^{1/2} \quad .
\end{equation}
The stall radius $R_{\rm f}$ in this case is only
1.3 times larger than for the adiabatic model, and the size
distribution follows the same law $N(R)\propto R^{1-2\beta}$ (Oey \&
Clarke 1997).  The observations of hot gas are therefore vital
confirmation that the adiabatic model applies to a significant
fraction of superbubbles. 

We can also derive the distribution of expansion
velocities $N(v)\ dv$, which describes only the
growing objects (Oey \& Clarke 1998b):
\begin{equation}
N_{\rm grow}(v) \propto v^{-7/2}\ , \quad \beta > 1.5 \quad .
\end{equation}
This again compares well with the SMC \hi\ shell catalog:  the
fitted power-law slope is $2.9\pm 1.4$.  Thus, despite the crude
assumptions in deriving the shell size and velocity distributions, the
data suggest that the neutral ISM in the SMC is fully consistent
with superbubble activity dominating the structure.  Although most
other available \hi\ shell catalogs are highly incomplete,
preliminary results for a few other galaxies also show agreement with
our model for the size distribution (Kim {\etal}1999; Mashchenko
{\etal}1999; Oey \& Clarke 1997).

\section{Correspondence with Star-Forming Regions}

One of the most obvious global tests of mechanical
feedback is to identify the parent stellar
populations, or their remains, with the superbubbles.  M31 (Brinks \&
Bajaja 1986) and M33 (Deul \& den Hartog 1990) both show 
correlations of OB assocations with \hi\ holes.  However,
Ho~II shows contradictory results, based on the \hi\
hole catalog compiled by Puche {\etal}(1992).  Tongue \& Westpfahl
(1995) found that the SN rate implied by radio continuum emission is
consistent with the hole energetics in that galaxy.  However, Rhode
{\etal}(1999) carried out a direct, $BVR$ search for remnant stellar
populations within the \hi\ holes, and found little evidence for the
expected stars.  But
Stewart {\etal}(2000) used far-UV data from the {\sl Ultraviolet
Imaging Telescope} and H$\alpha$ images to conclude that a significant
correlation between the \hi\ holes and recent star formation does
indeed support a feedback origin for the holes.

It is perhaps unsurprising that studies of Ho~II yield these confusing
results in view of that galaxy's distance of 3 Mpc.  The LMC, which
is 60 times closer, presents much better spatial resolution and
should therefore yield less ambiguous results.  Kim
{\etal}(1999) examined the correspondence between their \hi\
shell catalog, catalogued \hii\ regions (Davies {\etal}1976), and
H$\alpha$ imaging.  Not only do they find a correspondence, but they
are also able to identify an evolutionary sequence with respect to the
relative sizes, expansion velocities, and \Ha\ emission.  
Further investigation of the Magellanic Clouds should confirm and
reveal more quantitative details of the mechanical feedback 
process (Oey, Gerken, \& Walterbos, in preparation).  

\section{ISM porosity and galactic superwinds}

The consequences of feedback for the global ISM can be evaluated
quantitatively in terms of the interstellar porosity parameter $Q$,
which is the ratio:  (total area or volume occupied by
superbubbles) / (total area or volume of the galaxy).  Thus it is
essentially the filling factor of hot gas, assuming hot gas is contained
within all of the superbubbles.  Values of $Q$
near unity indicate the HIM dominates the multiphase ISM, and values
$\gg 1$ imply an outflow, with the galaxy generating more
hot gas than it can contain.

It is straightforward to use the analytic expression for $N(R)$
(equation~\ref{eqcaseC}) to derive $Q$ 
in terms of a galaxy's star-formation rate, $\Psi$ (Oey
{\etal}2001; see also Clarke \& Oey 2002):
\begin{equation}\label{eqQSFR}
Q \simeq 16 \frac{\Psi({\rm M_\odot\ yr^{-1}})}{hR_g^2(\rm kpc^3)}
	\qquad \propto \frac{1}{P_0} \ ,
\end{equation}
for $\beta=2$, a Salpeter (1955) IMF for stellar masses $0.1 \leq m
\leq 100\ \rm M_\odot$, and $P_0/k = 9500$.  $R_g$ and $h$ are the
radius of the gaseous star-forming disk and gas scale height, respectively.
We caution that $Q$ depends on ambient interstellar parameters, for
example, $P_0^{-1}$ as indicated. 

Oey {\etal}(2001) estimated $Q$ for all the galaxies in the Local
Group.  The Milky Way yields $Q\sim 1$ for some methods and $Q < 1$
for others, consistent with the ambiguous results found in the past
(e.g., McKee \& Ostriker 1977; Slavin \& Cox 1993).  The LMC
yields $Q\sim 1$, implying that hot gas dominates the ISM volume.  The
remainder of the Local Group  
galaxies all show $Q\ll 1$, with the sole exception of IC~10, a
starburst galaxy for which $Q\sim 20$, thereby unambiguously
predicting an outflow.  Oey {\etal}(2001) crudely estimate the
mass-loss rate in this outflow $\Mdot_{\rm out}$, assuming that the
material is largely evaporated from shell walls by thermal conduction.  
We find that $\Mdot_{\rm out} \sim \Psi$; indeed, absorption-line studies
of local starburst galaxies by Heckman {\etal}(2000) also show
that empirically, the outflow and star-formation rates have the same
order of magnitude for that sample.

Since $Q\sim 1$ represents a rough threshold for the escape of superwinds
from the galactic disk, this also implies the simultaneous escape of
newly-synthesized metals, which are contained in the hot gas.
Likewise, the shredding of the ISM into filaments facilitates the
escape of ionizing radiation, thus $Q\sim 1$ also represents an escape
threshold for ionizing photons (Clarke \& Oey 2002). 
We finally note the extensive body of numerical work on superbubbles and
blowout conditions.  Mac Low (1999) and Strickland \& Stevens (2000)
provide overviews of this field.  The details of the numerical
predictions are presently difficult to confirm empirically, but
observations with {\it XMM-Newton} and {\it Chandra} are beginning
to constrain the dominant processes. 

\section{Summary}

Observations of mechanical feedback ranging from
individual stellar wind bubbles to galactic superwinds are all
largely consistent with the conventional adiabatic model for shell
evolution.  Presently none of the discrepancies are of a magnitude
that suggest any need for major revision of the conventional
understanding.  The existence and properties of multiphase gas and
filamentary structure are broadly consistent with the adiabatic model.
It is also a remarkable strength that the model succeeds
across size scales ranging over at least three orders of magnitude.

However, characterizing the dominant parameters and their effects on
the shell evolution is still highly problematic.  For example,
critical ambient ISM conditions like density, pressure, and ionization
distributions remain elusive.  The mechanisms and conditions for
cooling of the interior energy need to be identified, and 
energy budgets reliably determined.  Perhaps the most
fundamental question is the fate of the hot gas generated within the
superbubbles:  Does it escape to constitute the HIM?  Does
it escape from starburst galaxies, and from their gravitational
potentials?  These issues have crucial consequences for 
galactic evolutionary processes, and our understanding depends on
further clarifying the mechanical feedback process.

\acknowledgements

I gratefully acknowledge support from the WS-ISM session organizers.


\theendnotes

\end{article}
\end{document}